\documentstyle[twoside,fleqn,espcrc2,psfig]{article}

\newcommand{\be}{\begin{equation}}
\newcommand{\ee}{\end{equation}}
\newcommand{\bea}{\begin{eqnarray}}
\newcommand{\eea}{\end{eqnarray}}

\title{Monte Carlo Renormalization Group calculation in $\lambda\phi^4_3$}
\author{Alex Travesset\address{Departament d'ECM, \\
                              Facultat de Fisica, \\
                              Universitat de Barcelona \\
                              Diagonal 647, 08027 Barcelona}
                     \thanks{Address after September 1st 1997,
			Department of Physics,
			         201 Physics Building,
                                 Syracuse,NY 13244-1130}}

\begin{document}

\begin{abstract}

We start by discussing some theoretical issues of renormalization
group transformations and Monte Carlo renormalization group technique.
A method to compute the anomalous dimension is proposed and investigated.
As an application, we find excellent values for critical exponents 
in $\lambda \phi^4_3$.
Some technical questions regarding the hybrid algorithm and 
strong coupling expansions, used to compute the critical couplings of the
canonical surface, are also briefly discussed.

\end{abstract}

\maketitle

\section{The MCRG calculation}

\subsection{Introduction}

The Monte Carlo renormalization group (MCRG) technique \cite{SWEN1},
consists of the
numerical determination of the couplings(and their derivatives) of
the RG transformed action. What is remarkable about this method 
is that, if the RG transformation(RGT) is sufficiently short ranged,
one may compute, in a finite lattice, 
critical quantities {\em as if working on an infinite lattice}.
	
Rather heuristically, one may 
see that as follows; suppose that a RGT possesses a Fixed Point(FP) 
consisting in a sum of local operators, each one just 
involving fields(or spins) 
separated at most by $n_s$ lattice sites. Assuming periodic
boundary conditions, this FP may be accommodated in a lattice as 
small as $(2n_s+1)^d$. Let us consider it now in a $(2(2n_s+1))^d$ volume, 
and apply a RGT, reducing it to a $(2n_s+1)^d$ lattice. As the
FP still fits in, one may expect the FP couplings (and their derivatives) 
not to feel the lattice (except
possibly by tiny effects dying off exponentially with the linear 
lattice size). Of course,
the same line of reasoning does not follow for observables, as it
is then the correlation length what should be fitted in, which is $\infty$.

It appears crucial then, both for a practical implementation 
and a theoretical understanding of the RG, to study the 
properties, such as locality of the FP, convergence of
eigenoperators, fast approach to the FP, etc. that 
different RGTs have. This is one of the reasons for this
project.

In this work, as we are dealing with unbounded spins, we must
determine the rescaling of the field, which at the FP is
related to the conformal anomalous dimension $\eta$. 
This is a difficult problem \cite{nos}, and we propose and investigate a
variation of the Bell-Wilson criteria \cite{BELL1}.


\subsection{The RGT}

RGTs in real space are (exponentially)short
ranged.The simplest of those, 
introduced by Bell and Wilson \cite{BELL1} is
\bea\label{Bell_transf}
e^{-S[\vartheta]}&=&\int \prod_{i=1}^{n}d \phi(n) \ 
e^{-S[\phi]} \times
\\\nonumber
&&\times \ e^{-\frac{a_W}{2}\sum_{n_B}(\vartheta(n_B)-b\sum_{n \in n_B} 
\phi(n))^2 } \ .
\eea
There are two free parameters $a_W$ and $b$. 

Recall that, for unbounded spins, we may rescale the fields.
Then, if a local FP exists, there is a whole line of equivalent FPs, so
there is a marginal (eigenvalue is 1) redundant direction. As we stick
to one of those FPs, we determine the parameter $b$ by preventing 
moves along this direction. Nevertheless, when one goes to a finite lattice
approximation, it is not clear that such a line of FPs still exists, 
the reason being that it is not guaranteed that all FPs are sufficiently
short ranged.  In the other case, 
the parameter $a_W$ just labels different RG transformations. It is
a free parameter at our disposal to play with.

\begin{figure}[htb]
\vspace{9pt}
\centerline{\psfig{file=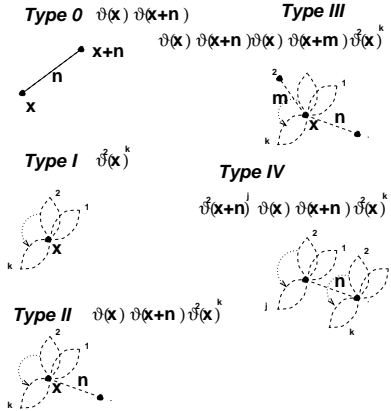,width=5.2cm}}
\caption{Operators included in the RG transformed action.}
\label{fig__OP1}
\end{figure}

Our canonical surface is
\bea\label{cano_surf}
S[\phi]&=&\sum_n \frac{1}{2}\sum_{\mu} (\phi(n)-\phi(n+\mu))^2+
\\\nonumber
&&+\frac{m^2}{2}\phi(n)^2+\frac{\lambda}{4} \phi(n)^4 \ ,
\eea
and we expand the RG transformed action in Eq.~\ref{Bell_transf}
as a sum
\be\label{ansatz_trans}
S[\vartheta]=\sum_i O_i(\vartheta) \ ,
\ee
where the set of local operators are shown in fig.~\ref{fig__OP1}.

\section{The critical surface}

Following the discussion in the preceding section,
a MCRG calculation must be done at the exact critical point of the
infinite volume system. In Eq.~\ref{cano_surf},
for each value of $\lambda$ there exists a critical value
$m_c(\lambda)$. A strong coupling expansion, which
for this model amounts to an expansion in the hopping parameter $\kappa$,
allows to extract all critical couplings with negligible CPU time.

The first 11 coefficients of this expansion were tabulated in \cite{STRONG}.
To extract critical couplings from those,
presents some technical difficulties related to the appearance of
antiferromagnetic singularities in loose-packed lattices, which
nevertheless, may be overcome.  Critical couplings 
used are given in table \ref{tab__CS}.

  \begin{table}[t]
  \centerline{
  \begin{tabular}{||l|l||l|r||}
  \multicolumn{1}{c}{$g$} & \multicolumn{1}{c}{$\kappa$} &
  \multicolumn{1}{c}{$m^2$} & \multicolumn{1}{c}{$\lambda$}   \\\hline
   0.5    & 0.3973(1) & -6.0000(30) & 12.672(6) \\\hline
   0.45   & 0.3976(1) & -5.4969(26) & 11.388(6) \\\hline
   0.39458& 0.3973(1) & -4.9386(25) & 10.000(6) \\\hline
   0.3    & 0.3948(1) & -3.9736(21) &  7.700(6) \\\hline
   0.2287 & 0.3905(2) & -3.2207(20) &  6.000(5) \\\hline
   0.15204& 0.3820(2) & -2.3568(14) &  4.167(3) \\\hline
   0.005  & 0.3369(-) & -0.1230(--) &  0.176(-) \\\hline
   0.00   & 0.3333(-) & -0.0000(--) &  0.000(-) \\\hline
  \end{tabular}} 
  \caption{Critical couplings at the region of interest,
   if error bars are not written, all digits are significant.}
  \label{tab__CS}
  \end{table}

\section{The algorithm}

 The algorithm we used in our simulation is
 the Hybrid algorithm \cite{HYB}, which has two freedoms, 
 namely, the size of the leap frog step $\delta t$, 
 and the number of integration
 steps $n_l$ before the Metropolis test is passed. While there is not
 much room to play with $\delta t$, big gains come in playing 
 with $n_l$.

 Indeed, in table \ref{tab__auto} is reported a sample calculation.
 Gradually increasing $n_l$, autocorrelations
 reduce, with a minor penalty in CPU time. In a 
 $24^3$ lattice, we took $n_l=60$ or $n_l=120$.

  \begin{table}[b]
  \centerline{
  \begin{tabular}{|c|c||l|r|r|}
  \multicolumn{1}{c}{$n_l$} & \multicolumn{1}{c}{$\delta t$} &
  \multicolumn{1}{c}{$\langle O_1 \rangle $} & 
  \multicolumn{1}{c}{$\tau_{O_1}$} & \multicolumn{1}{c}{$CPU$} 
                       \\\hline
  6  & 0.01 & 0.27449(140) &  500 & 140    \\\hline
  6  & 0.03 & 0.27425( 57) &  215 & 140    \\\hline
  6  & 0.06 & 0.27506( 40) &  100 & 140    \\\hline
  11 & 0.06 & 0.27527( 23) &   35 & 160    \\\hline
  21 & 0.06 & 0.27510( 12) &    8 & 189    \\\hline
  31 & 0.06 & 0.27511(  6) &    4 & 218    \\\hline
  \end{tabular}  
  } 
  \caption{The results correspond to a $\lambda \phi^4_3$ theory in a
  $10^3$ lattice at $m^2=-3.32$,
  $\lambda=6.0$, $1.2\times10^6$ configurations, throwing
  away $6\times10^5$ for thermalization, $O_1=\phi^2$, 
   $\tau$ is the autocorrelation
  time, $CPU$ is in units of $10^3$ sec.~in a Power 
  Challenge.}
   \label{tab__auto}
   \end{table}

\section{Results}

We perform 3 RGTs for different $a_W$ values in a $24^3$ lattice
including up to 27 operators. For each value of $a_W$, 
there exists a $b$ such that
the eigenvalue of the T matrix in the second RGT is $1$(we discard
the first RGT as we assume we are not yet on the linear region around
the FP). If this value for $b$ makes the marginal eigenvalue stable
against the last RGT, we assume we reached a local FP. So, we fix $b$ and
choose the $a_W$ so that the canonical surface, Eq.~\ref{cano_surf},
is optimally close to the FP. There are two cross checks.
First, computing the flow (using of Schwinger-Dyson
equations \cite{GONZ}), and secondly, performing 2 RGTs 
in a $12^3$ lattice, and testing if expectation values of operators in 
the last RGT agree
with the ones coming from the third transformation on a $24^3$ 
lattice.  

For small values of $\lambda$, though strictly we are in the 
domain of the Wilson FP,
we are on the linear region around the Gaussian FP, and indeed, results
are totally compatible with those.

\begin{figure}[htb]
\vspace{9pt}
\centerline{\psfig{file=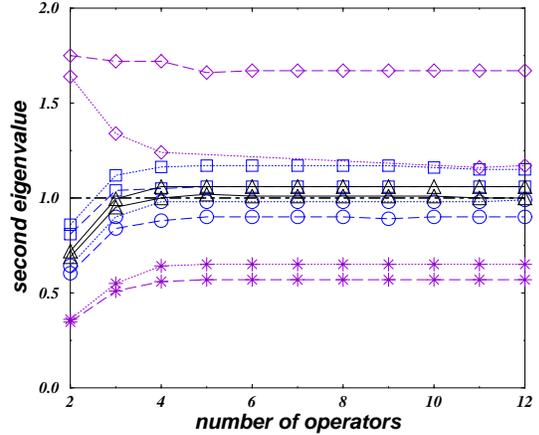,width=7cm}}
\caption{Second eigenvalue at $\eta=0.03$. diamonds correspond to 
$a_W=\infty$, squares $a_W=35$, triangles $a_W=25$, circles $a_W=20$
and asterisks $a_W=8$. Dotted line is the 2nd RGT and 
dashed line the 3rd one.}
\label{fig__EIG}
\end{figure}

At the strong coupling we studied different values of $\lambda$.
The most extensive one was performed at $\lambda=12.672$, where
the optimal values where $a_W=25$ and $\eta=0.03$. As an example,
in fig.~\ref{fig__EIG} we plot the value of the second eigenvalue
at $\eta=0.03$ for different values of $a_W$ clearly  
singling out $a_W=25$. The couplings of the action in the last two RGTs 
agree within statistical error bars, though those are surprisingly large.
Furthermore, the agreement
between expectation of the third RGT and the ones that come from a
second RGT starting in a $12^3$ differ, at most, by a $5 \%$,
which is good but not fully satisfactory yet.

Our preliminary results for the critical exponents are 
\bea\label{crit_expo}
\eta=0.030(5) \ &,& \ \nu=0.625(7) \  
\\\nonumber
\ \omega=0.77(5) \ \ &,& \ \lambda_{-2}=0.28(9) \ ,
\eea
where error bars are just tentative. Those figures
are excellent when compared with the most accepted results
$\eta \sim 0.035 \ , \ \nu \sim 0.631 \ , \ \omega \sim 0.8$,
(2th irrelevant eigenvalue is not known to us).

\section{Acknowledgments}

It is a pleasure to acknowledge interest and discussions with B.Alles,
S. Catterall,
J.Comellas, D.Espriu and R. Toral. I have benefited from a fellowship
of the Generalitat of Catalunya. Part of the numerical work was done at
C.E.P.B.A.This work has been supported by
grants AEN95-0590 (CYCIT) and GRQ93-1047(CIRIT).

\end{document}